\PassOptionsToPackage{hyphens}{url}
\PassOptionsToPackage{dvipsnames,svgnames,x11names}{xcolor}
\documentclass[
  12pt]{article}
 
\RequirePackage{amsthm,amsmath,amsfonts,amssymb} 
\RequirePackage[authoryear]{natbib} 
\RequirePackage[authoryear]{natbib}
\usepackage{xr-hyper}  
\RequirePackage[colorlinks=true, 
    citecolor=blue, 
    urlcolor=red, 
    linkcolor=red,    
    filecolor=red  
    ]{hyperref}
\usepackage{multirow}
\usepackage{algorithmic} 
\RequirePackage{graphicx}
\usepackage{amsmath}
\usepackage{bm}
\usepackage{graphicx} 
\usepackage{natbib}
\usepackage{url}  
\usepackage{booktabs}
\usepackage{multirow}
\usepackage{algorithm} 
\usepackage{enumerate}
\usepackage{natbib}
\usepackage{tabularx} 
\usepackage{lscape} 
\usepackage{url} 
\usepackage{enumitem} 
\usepackage{multirow}
\usepackage{accents}
\usepackage{graphicx}
\usepackage{rotating}
\usepackage{subcaption}
\usepackage{mwe}
\usepackage{float}
\usepackage{booktabs,makecell}
\usepackage{amsthm,amssymb}
\usepackage{amsfonts} 
\usepackage{dsfont}
\usepackage{amsmath}
\usepackage{amssymb}
\usepackage{booktabs}
\usepackage{multirow} 
\usepackage{multirow}
\usepackage{booktabs}
\usepackage{bbm}
\usepackage{mathrsfs} 
\usepackage{bbm}
\usepackage{booktabs}
\usepackage{tcolorbox}
\usepackage{multirow}
\usepackage{rotating} 
\usepackage[normalem]{ulem}
\usepackage{tablefootnote}  
\usepackage{threeparttable}

\usepackage[normalem]{ulem}
\newcommand{\ul}[1]{\uline{#1}}

\definecolor{darkgreen}{RGB}{0, 176, 80}
\definecolor{islrorange}{RGB}{205, 100, 32}
\definecolor{islrblue}{RGB}{93, 165, 216}
\definecolor{whaleblue}{RGB}{51, 51, 178}
\definecolor{highlightblue}{RGB}{68, 68, 255}
\definecolor{pptorange}{RGB}{237, 125, 49}
\definecolor{pptgreen}{RGB}{112, 173, 71}
\definecolor{pptpurple}{RGB}{114, 51, 162}
\definecolor{highlightred}{RGB}{204, 0, 0}

\newcommand{\R}{\mathbb{R}}
\newcommand{\N}{\mathbb{N}}

\newtheorem{theorem}{Theorem}
\newtheorem{remark}{Remark}

\newtheorem{corollary}{Corollary}
\newtheorem{assumption}{Assumption}

\theoremstyle{plain}

\theoremstyle{definition}
\newtheorem{definition}{Definition}

\usepackage{amsmath,amssymb}
\usepackage{iftex}
\ifPDFTeX
  \usepackage[T1]{fontenc}
  \usepackage[utf8]{inputenc}
  \usepackage{textcomp} 
\else 
  \usepackage{unicode-math}
  \defaultfontfeatures{Scale=MatchLowercase}
  \defaultfontfeatures[\rmfamily]{Ligatures=TeX,Scale=1}
\fi
\usepackage{lmodern}
\ifPDFTeX\else  
\fi
\IfFileExists{upquote.sty}{\usepackage{upquote}}{}
\IfFileExists{microtype.sty}{
  \usepackage[]{microtype}
  \UseMicrotypeSet[protrusion]{basicmath} 
}{}
\makeatletter
\@ifundefined{KOMAClassName}{
  \IfFileExists{parskip.sty}{
    \usepackage{parskip}
  }{
    \setlength{\parindent}{0pt}
    \setlength{\parskip}{6pt plus 2pt minus 1pt}}
}{ 
  \KOMAoptions{parskip=half}}
\makeatother
\usepackage{xcolor}
\makeatletter
\ifx\paragraph\undefined\else
  \let\oldparagraph\paragraph
  \renewcommand{\paragraph}{
    \@ifstar
      \xxxParagraphStar
      \xxxParagraphNoStar
  }
  \newcommand{\xxxParagraphStar}[1]{\oldparagraph*{#1}\mbox{}}
  \newcommand{\xxxParagraphNoStar}[1]{\oldparagraph{#1}\mbox{}}
\fi
\ifx\subparagraph\undefined\else
  \let\oldsubparagraph\subparagraph
  \renewcommand{\subparagraph}{
    \@ifstar
      \xxxSubParagraphStar
      \xxxSubParagraphNoStar
  }
  \newcommand{\xxxSubParagraphStar}[1]{\oldsubparagraph*{#1}\mbox{}}
  \newcommand{\xxxSubParagraphNoStar}[1]{\oldsubparagraph{#1}\mbox{}}
\fi
\makeatother

\usepackage{longtable,booktabs,array}
\usepackage{calc} 
\usepackage{etoolbox}
\makeatletter
\patchcmd\longtable{\par}{\if@noskipsec\mbox{}\fi\par}{}{}
\makeatother
\IfFileExists{footnotehyper.sty}{\usepackage{footnotehyper}}{\usepackage{footnote}}
\makesavenoteenv{longtable}
\usepackage{graphicx}
\makeatletter
\def\maxwidth{\ifdim\Gin@nat@width>\linewidth\linewidth\else\Gin@nat@width\fi}
\def\maxheight{\ifdim\Gin@nat@height>\textheight\textheight\else\Gin@nat@height\fi}
\makeatother
\setkeys{Gin}{width=\maxwidth,height=\maxheight,keepaspectratio}
\makeatletter
\def\fps@figure{htbp}
\makeatother

\makeatletter
\@ifpackageloaded{caption}{}{\usepackage{caption}}
\AtBeginDocument{%
\ifdefined\contentsname
  \renewcommand*\contentsname{Table of contents}
\else
  \newcommand\contentsname{Table of contents}
\fi
\ifdefined\listfigurename
  \renewcommand*\listfigurename{List of Figures}
\else
  \newcommand\listfigurename{List of Figures}
\fi
\ifdefined\listtablename
  \renewcommand*\listtablename{List of Tables}
\else
  \newcommand\listtablename{List of Tables}
\fi
\ifdefined\figurename
  \renewcommand*\figurename{Figure}
\else
  \newcommand\figurename{Figure}
\fi
\ifdefined\tablename
  \renewcommand*\tablename{Table}
\else
  \newcommand\tablename{Table}
\fi
}
\@ifpackageloaded{float}{}{\usepackage{float}}
\floatstyle{ruled}
\@ifundefined{c@chapter}{\newfloat{codelisting}{h}{lop}}{\newfloat{codelisting}{h}{lop}[chapter]}
\floatname{codelisting}{Listing}

\makeatother
\makeatletter
\@ifpackageloaded{caption}{}{\usepackage{caption}}
\@ifpackageloaded{subcaption}{}{\usepackage{subcaption}}
\makeatother

\ifLuaTeX
  \usepackage{selnolig}  
\fi
\usepackage[]{natbib}
\usepackage{bookmark}

\IfFileExists{xurl.sty}{\usepackage{xurl}}{}

\newcommand{\anon}{1}

\def\spacingset#1{\renewcommand{\baselinestretch}
{#1}\small\normalsize} \spacingset{1}

\begin{document}

\if1\anon
{
  \title{\bf  
  NetworkNet: A Deep Neural Network Approach for Random Networks with Sparse Nodal Attributes and Complex Nodal Heterogeneity}
  \author{Zhaoyu Xing \hspace{.2cm}\\
    \small{ Department of Applied and Computational Mathematics and Statistics} \\ 
\small{ University of Notre Dame,} \\
    \small{and} \\
    Xiufan Yu \\
     \small{ Department of Applied and Computational Mathematics and Statistics}\\
\small{ University of Notre Dame} }
  \maketitle
} \fi

\if0\anon
{
  \bigskip
  \bigskip
  \bigskip
  \begin{center}
    {\LARGE\bf Title}
\end{center}
  \medskip
} \fi


\begin{abstract}
Heterogeneous network data with rich nodal information become increasingly prevalent across multidisciplinary research, 
yet accurately modeling complex nodal heterogeneity and simultaneously selecting influential nodal attributes remains an open challenge. 
This problem is central to many applications in economics and sociology, when both nodal heterogeneity and high-dimensional individual characteristics highly affect network formation. 
We propose a statistically grounded, unified deep neural network approach for modeling nodal heterogeneity in random networks with high-dimensional nodal attributes, namely ``NetworkNet''.  
A key innovation of NetworkNet lies in a tailored neural architecture that explicitly parameterizes attribute-driven heterogeneity, and at the same time, embeds a scalable attribute selection mechanism. NetworkNet consistently estimates two types of latent heterogeneity functions, i.e., nodal expansiveness and popularity, while simultaneously performing data-driven attribute selection to extract influential nodal attributes. By unifying classical statistical network modeling with deep learning, NetworkNet delivers the expressive power of DNNs with methodological interpretability, algorithmic scalability, and statistical rigor with a non-asymptotic approximation error bound.  
Empirically, simulations demonstrate strong performance in both heterogeneity estimation and high-dimensional attribute selection. 
We further apply NetworkNet to a large-scale author-citation network among statisticians, revealing new insights into the dynamic evolution of research fields and scholarly impact. 
\end{abstract}

\noindent%
\vfill

\newpage
\spacingset{1.8}

\section{Introduction} 
Sociological and economic entities frequently interact through complex networks. Within these structures, firms trade goods, banks extend credit, and researchers disseminate knowledge through citations. A key characteristic of these networks is the underlying heterogeneity of the individuals (e.g., firm size, risk profile, research expertise) that shape both how actively they initiate interactions and how strongly they attract connections from others. Quantifying this individual-level heterogeneity is therefore a crucial step in modeling network formation, evaluating systemic evolutions, and designing effective policies.

In practice, network data with rich nodal information are ubiquitous in modern science. Alongside rapidly expanding network scale, modern network datasets exhibit greater structural complexity and growing granularity in edge information, evolving from simple binary undirected networks to weighted directed networks.
Meanwhile, nodes themselves are often accompanied by rich attributes describing the individual entities, as seen in online social networks with user-specific features \citep{leskovec2012learning, xing2025regularization}, academic collaboration networks annotated with scholars' research interests \citep{gao2023large, gao2025JMLRaccepted}, economic networks with different individual characteristics \citep{ANTINYAN2020103547, xing2024palms, xing2025calms}, and among others.  
Classical random network models primarily focus on the topological structure and treat all nodes as statistically equivalent \citep{ERnetwork1959, watts1998Nature, schweinberger2011instability,chatterjee2013AOS}. In practice, however, real-world networks are formed among intrinsically heterogeneous entities, where influential node-level attributes can substantially impact the generative mechanisms of network connections. 

Modeling nodal heterogeneity, driven by the practical needs in network analysis, has thus become a critical yet challenging problem in network science. For example, in academic social networks, established researchers typically exert greater collaborative and academic influence than early-career researchers who often possess fewer contributions concentrated in limited research areas. Overlooking such inherent individual nodal heterogeneity risks oversimplifying the underlying connection mechanisms and may lead to biased assessments of scholarly impact.  

As a classic application in social networks, academic collaboration and scholarly impact have long been central topics. Nevertheless, previous findings on scholarly impact analysis are solely based on network structures, and do not fully account for the effect of nodal heterogeneity \citep{lehmann2003citation, moody2004structure, che2025flat}. Although some approaches attempt to capture nodal heterogeneity using local topological structures \citep{ma2025supervised, Yan2016AOS}, they rely exclusively on the network structures and are unable to fully utilize the high-dimensional nodal attributes now widely available in modern empirical data.

The availability of rich nodal information in modern academic network data opens new opportunities for characterizing heterogeneity beyond network topology alone. 
A large academic network dataset \citep{gao2023large} was recently released, containing count-valued edges and high-dimensional nodal attributes for 168,171 scholars 
based on 97,436 papers in representative statistical journals over a four-decade span. 
This dataset enables empirical investigations of attribute-driven nodal heterogeneity within an edge-generative random network, and provides a unique basis for studying how individual-level attributes affect the scholarly impact through heterogeneous network generative processes.

This paper is empirically motivated by both the practical need for principled quantification of scholarly impact using this newly released public dataset of academic networks with informative count-valued edges and rich nodal attributes, and the lack of methodologies for  modeling nodal heterogeneity with high-dimensional nodal attributes. We seek to fill  this methodological gap and address the following fundamental question:

\textit{How to better quantify the nodal heterogeneity with high-dimensional nodal attributes?} 
Prior works in network modeling reveal a clear methodological divide between statistics and deep learning perspectives:
statistical network models principally focus on characterizing randomness in network topology without modeling the attribute-driven nodal heterogeneity, while deep learning approaches typically analyze observed networks as fixed input structures without considering the nodal heterogeneity in the generating process of random networks. To our knowledge, no method yet adequately unifies the strengths of statistical and deep learning techniques for modeling nodal heterogeneity using high-dimensional nodal attributes. 

Specifically, related statistical network models treat the networks as random variables, and largely focus on the generating processes of network structures as well as the corresponding network distributions. Among them, the 
exponential random graph models (ERGMs) capture structural randomness through their topology factors \citep{hunter2008ergm,reviewERGM}. However, ERGMs are computationally intensive and face inferential challenges, which become even more demanding for extensions to count-valued networks \citep{krivitsky2012exponential,HB2024}, and their classical formulations assume node exchangeability that limits their ability to model nodal heterogeneity or incorporate influential node-specific variation. %
Stochastic block models capture the random interactions in a network with estimated community information of nodes \citep{LeiRinaldo,lei2023bias, zhang2023adjusted}. Although they allow the different probabilities for connection among different communities, they usually fail to capture the individual nodal heterogeneity as the nodes within one community are often assumed to be identical. 
Degree-based models, such as the beta-model and its variants \citep{Yan2016AOS, graham2017econometric, stein2025sparse},  utilize the degree heterogeneity to capture the randomness of network structures where latent parameters are estimated to describe the nodal heterogeneity.
However, previously proposed classic beta-models focus on degrees of binary networks and model the network structures only. While few recent extensions can incorporate covariates to capture the nodal heterogeneity \citep{wang2024variable, Yan2019JASA}, they are specially designed for edge-wise features and assume restrictive linear relationships. 
Commonly collected nodal attributes are not directly used to capture the nodal heterogeneity in random network models. Despite the clear practical need, a general statistical method that can model complex count-valued networks with high-dimensional nodal attributes and non-linear effects 
remains an open challenge.

Parallel to statistical network models, 
deep learning fields have recently made significant strides in network analysis, particularly in tasks such as link prediction and node classification \citep{AnnieQu2023high, JMLR2018Review}. 
However, almost all of these related methods treat networks as fixed given structures that serve as inputs of the algorithms without considering the randomness in network generating process. A growing body of work has shown that the neural network approaches for network data can be highly sensitive to random noise and attacks in data \citep{zugner2018adversarial, wang2024uncertainty,goodfellow2014explaining, szegedy2013intriguing}. Overlooking the inherent stochasticity of networks is one of the main reasons for overfitting problems while modeling network data. 
Besides, most deep learning-based graph models function as black boxes that offer limited interpretability within the theoretical context of network science, despite their strong predictive performance in practice.
While some ``explainer'' models are proposed for graph neural networks \citep{vu2020pgm, ying2019gnnexplainer}, they focus mainly on estimating the feature importance for specific learning tasks but do not evaluate their influence on network distribution. Thus, without accounting for the probabilistic generating nature of random networks, these deep learning-only approaches are unable to effectively capture and interpret the nodal heterogeneity.

Despite rapid methodological advances in statistics and deep learning techniques for network data, there remains a lack of methods capable of explicitly modeling nodal heterogeneity using high-dimensional nodal attributes in random networks, leaving a gap in the literature. To close this critical gap, we propose a novel framework for modeling nodal heterogeneity in count-valued directed random networks with high-dimensional nodal attributes, namely ``NetworkNet''. 
NetworkNet is a powerful unified approach that allows us to consistently model the nodal heterogeneity utilizing both large-scale count-valued networks and high-dimensional nodal information, and effectively identify the influential subset of nodal attributes.
As a synthesis of statistical and deep learning methods, NetworkNet models the count-valued random edges with Poisson distributions, where the heterogeneous rates for edge generation are determined by both the sender's and receiver's nodal heterogeneity. Influential nodal attributes are selected from the set of high-dimensional nodal attributes and used to capture complex nodal heterogeneity through specially designed deep neural networks (DNN). 
For the architectures of NetworkNet, we introduce two skip layers with corresponding constraints based on hierarchical $L_1$ regularization to accurately estimate the nodal heterogeneity and select nodal attributes simultaneously. Theoretically, we show that the NetworkNet estimator can approximate the latent nodal heterogeneity functions consistently and achieve the non-asymptotic upper bound of the error. 

This work contributes to the network literature by addressing several critical challenges in a unified framework. First, we introduce a nodal attribute selection mechanism designed for the network context, enabling us to sift through high-dimensional nodal information and identify meaningful attributes. Second, we propose a principled statistical model for count-valued networks that directly targets the distribution of edge counts, effectively accommodating nodal heterogeneity in the interaction intensity. Third and most notably, we synthesize statistical methods with modern deep learning techniques by embedding flexible neural components within a coherent probabilistic framework, bridging statistical rigor and deep learning expressiveness. This allows us to leverage the representational power of DNNs for complex nodal information while retaining a statistical-model-based treatment of edge counts and a clear link between nodal attributes and edge intensities. Together, NetworkNet offers a powerful tool that is simultaneously distributionally appropriate for count-valued edges, scalable to high-dimensional nodal attributes, and capable of integrating predictive flexibility with interpretable statistical learning. Furthermore, NetworkNet is not limited to count-valued network data, but offers a flexible and unified modeling framework that can be readily extended to capture nodal heterogeneity in a broad class of network types, such as undirected networks, binary networks, and continuously weighted networks. Empirically, the proposed general approach NetworkNet can be applied broadly to analyze not only academic collaboration networks, and also other practical network structures with nodal attributes, such as the social networks with personal characteristics, business networks with companies' features and international trading networks with macroeconomic indexes.

The rest of this paper is organized as follows. Section 2 formally introduces the NetworkNet model to estimate the nodal heterogeneity and select relevant nodal attributes with an alternating optimization algorithm. We demonstrate the dominance of NetworkNet in both nodal heterogeneity estimation and nodal attribute selection with comprehensive simulation studies in Section 3. In Section 4, we utilize the proposed NetworkNet to conduct an in-depth analysis of the academic author-citation networks over a four-decade span, investigating the evolution of scholarly impact and uncovering data-driven insights. Section 5 concludes with a discussion. Proofs of all theoretical results and additional details of empirical analysis are included in the Supplementary Materials.

\section{Methods} 

To introduce our approach to model nodal heterogeneity in random networks with high-dimensional nodal attributes clearly, we first introduce the necessary notation and basic model settings. Then we introduce the framework of NetworkNet, followed by the iterative optimization algorithms. Subsequently, theoretical results, including a non-asymptotic bound, are provided.

\subsection{Notation} 

We consider a count-valued directed network with $n$ vertices and corresponding adjacency matrix $\bm{A} = (A_{ij})_{n \times n}$, $i,j \in [n]$, where $A_{ij}\in \mathbb{N}$ represents the count-valued weight of the edge between the nodes $i$ and $j$, and $[n]=\{1,\cdots,n\}$. For any constant $a\in\mathbb{R}$, $\lfloor a \rfloor$ represents the largest integer that is smaller than or equal to $a$. Besides the network structures, a $p$-dimensional vector of nodal covariates associated with each node is denoted by $\bm{x}_i \in \mathbb{R}^p$, $i\in[n]$. These covariates may include both continuous and categorical attributes. The complete set of nodal attributes for the network is represented by the matrix $\bm{X} = [\bm{x}_1, \bm{x}_2, \dots, \bm{x}_n]' \in \mathbb{R}^{n \times p}$, which characterizes the nodal heterogeneity. 
For a function $\eta(\bm{x}_i) \in \mathcal{F}, i\in [n]$, the $L_{n,2}$ norm defined on class $\mathcal{F}$ is $\|\eta(\cdot)\|_{n,2}=[\sum_{i=1}^n\eta(\bm{x}_i)^2/n]^{1/2}$. 
For any compact space $\mathcal{X}\subseteq [a,b]^p$ with reference measure $\mu$, define $
\|\eta\|_{L_2} = \left[\int_{\mathcal{X}} \eta(\bm{x})^2\,d\mu(\bm{x})\right]^{1/2}$ for a measurable function $\eta:\R^p\to\R$. For the pair $\eta=(\alpha,\beta)\in\mathcal{G}(\mathcal{F})$ from a function class $\mathcal{G}(\mathcal{F})$, define $\|\eta\|_{L_2} = \left(\|\alpha\|_{L_2}^2+\|\beta\|_{L_2}^2\right)^{1/2}.$
$\lceil r \rceil$ represents a ceiling function that indicates the smallest integer that is greater than or equal to $r$. For a vector $\bm{\theta}=(\theta_1,\cdots,\theta_n)$, $\|\bm{\theta}\|_1=\sum_{i=1}^n|\theta_i|$ represents the $L_1$-norm of the vector $\bm{\theta}$. We denote scalars and functions using lowercase letters, and denote matrices and vectors by bold uppercase and bold lowercase letters, respectively.

\subsection{NetworkNet} 
  
Motivated by the author-citation networks, where citation interactions are initiated at irregular time points and citation counts represent the aggregation of these discrete random events, we model the count-valued network data via a set of Poisson distributions with latent parameters as 
\begin{equation}
    A_{ij} | \lambda_{ij} \sim \mathrm{Poisson}(\lambda_{ij}),
    \label{ModelSetting}
\end{equation}
where the non-negative rate parameter $\lambda_{ij}$ represents the expected number of interactions from node $i$ to node $j$ within a time range. For academic author-citation networks, the number of citations coming from different publications can only be count-valued data, which is modeled by \eqref{ModelSetting} with Poisson distribution, and the general strength of the connection between scholars is captured by the Poisson parameter $\lambda$, which is assumed to be determined by both the \textit{expansiveness} of origin node $i$ and the \textit{popularity} of the destination node $j$.  

The high-dimensional nodal attributes give a straightforward description about individual characteristics that are usually strongly related to the network generating process. To better model the nodal heterogeneity in the generating process of random graphs, we introduce two types of latent node-specific components based on high-dimensional nodal attributes, the \textit{expansiveness function} $\alpha(\cdot)$ capturing the nodal inherent heterogeneity in generating connections to the other nodes, and the \textit{popularity function} $\beta(\cdot)$ capturing the nodal inherent heterogeneity in attracting  connections from the other nodes. Both of them are functions of nodal attributes from a general function class and capture two types of the nodal heterogeneity. Then we model the rate $\lambda_{ij}$ of the count-valued edges $A_{ij}$ in \eqref{ModelSetting} with both the \textit{expansiveness} of the edge-sender and \textit{popularity} of the edge-receiver as 
\begin{equation}
    \log(\lambda_{ij}) = \left[ \alpha(\bm{x}_i) + \beta(\bm{x}_j) \right]/ Z_n,
    \label{eq:model_main}
\end{equation}
where $Z_n$ is a scaling constant that ensures that expected edge counts 
remain well-behaved as the network grows, analogous to the 
sparsity-inducing scaling commonly adopted in large-network 
asymptotics. As a constant shift can lead to model identifiability issues, we set $\sum_{i=1}^n\alpha(\bm{x}_i)=\sum_{j=1}^n\beta(\bm{x}_j)$, while other constraints can also be used.
 
The flexible expansiveness function $\alpha(\cdot)$ and popularity function $\beta(\cdot)$ in our model settings can capture complex non-linear relationships between nodal attributes and nodal heterogeneity that is directly related to the probability distribution of random networks. Modeling heterogeneity with high-dimensional and sparse nodal attributes in random network models is an open challenge. To address this problem, we embed flexible neural techniques within the coherent probabilistic framework to model the nodal heterogeneity in large-scale random networks, and select the high-dimensional nodal attributes simultaneously. Specifically, we approximate latent nodal heterogeneity functions $\alpha(\cdot)$ and $\beta(\cdot)$ with deep neural networks from the following class     
\begin{equation}
\label{ClassH}
    \mathcal{H}=\{f \equiv f(\bm{x},\bm{W},\bm{\theta}): \bm{x} \mapsto \bm{\theta}'\bm{x} + h_{\bm{W}}(\bm{x}) \} , 
\end{equation} 
where $\bm{\theta}$ represents the coefficient vector for the skip layer and $h_{\bm{W}}(\cdot)$ represents a ReLU-DNN with weights $\bm{W}$. Both $\bm{W}$ and $\bm{\theta}$ are parameters to be estimated. To select the influential nodal attributes while modeling nodal heterogeneity within a probabilistic framework, we formulate a regularized optimization problem minimizing a composite objective function with the negative log-likelihood of the network under the Poisson assumption, additional regularization terms and constraints:
\begin{align}
&\min_{\bm{\theta}_\alpha,\bm{\theta}_\beta, \bm{W}_\alpha,\bm{W}_\beta}  \sum_{i \neq j} \mathcal{L}_{ij}(\bm{\theta}_\alpha,\bm{\theta}_\beta,\bm{W}_\alpha,\bm{W}_\beta)
    + \lambda_1  \|\bm{\theta}_\alpha\|_1 + \lambda_2  \|\bm{\theta}_\beta\|_1   , \label{eq:optimization_problem} \\
     & s.t. ~ \|W_{\alpha,j}^{(1)}\|_\infty\leq M |\theta_{\alpha,j}| , \|W_{\beta,j}^{(1)}\|_\infty\leq M |\theta_{\beta,j}|
    \text{ and }  \sum_{i=1}^n\alpha(\bm{x}_i)=\sum_{j=1}^n\beta(\bm{x}_j),  \nonumber 
\end{align}
where two neural networks $f(\bm{X}_i;\bm{W}_{\alpha},\bm{\theta}_\alpha)$ and $g(\bm{X}_j;\bm{W}_{\beta},\bm{\theta}_\beta)$ with skip-layers are used to approximate 
latent nodal heterogeneity functions $\alpha(\cdot)$ and $\beta(\cdot)$,
and the empirical loss function is 
\begin{align}
    \mathcal{L}_{ij}(\bm{\theta}_\alpha,\bm{\theta}_\beta,\bm{W}_\alpha,\bm{W}_\beta) =   \exp \left[f(\bm{x}_i;\bm{W}_{\alpha},\bm{\theta}_\alpha)/Z_n + g(\bm{x}_j;\bm{W}_{\beta},\bm{\theta}_\beta)/Z_n\right]  \nonumber \\
- A_{ij} \left[f(\bm{x}_i;\bm{W}_{\alpha},\bm{\theta}_\alpha)/Z_n + g(\bm{x}_j;\bm{W}_\beta,\bm{\theta}_\beta)/Z_n\right]  \label{empiricalLoss}
\end{align}
with observed network $\bm{A}=\{A_{ij}\}$ and nodal attributes $\bm{X}$. The $L_1$ penalty is applied to shrink the weights of the skip layers with tuning-parameters $\lambda_1, \lambda_2>0$, and $\bm{\theta}_\alpha$ and $\bm{\theta}_\beta$ serve as the constraints for the first-layer weights $\bm{W}_{\alpha}^{(1)}$ and $\bm{W}_{\beta}^{(1)}$ in DNNs. When $(\hat{\bm{\theta}}_\alpha)_{j}=0$, then the corresponding feature $j$ is effectively removed from the subset of influential nodal attributes for nodal expansiveness, thus achieving data-driven nodal attributes selection. 

\begin{figure}
    \centering
    \includegraphics[width=0.7\linewidth]{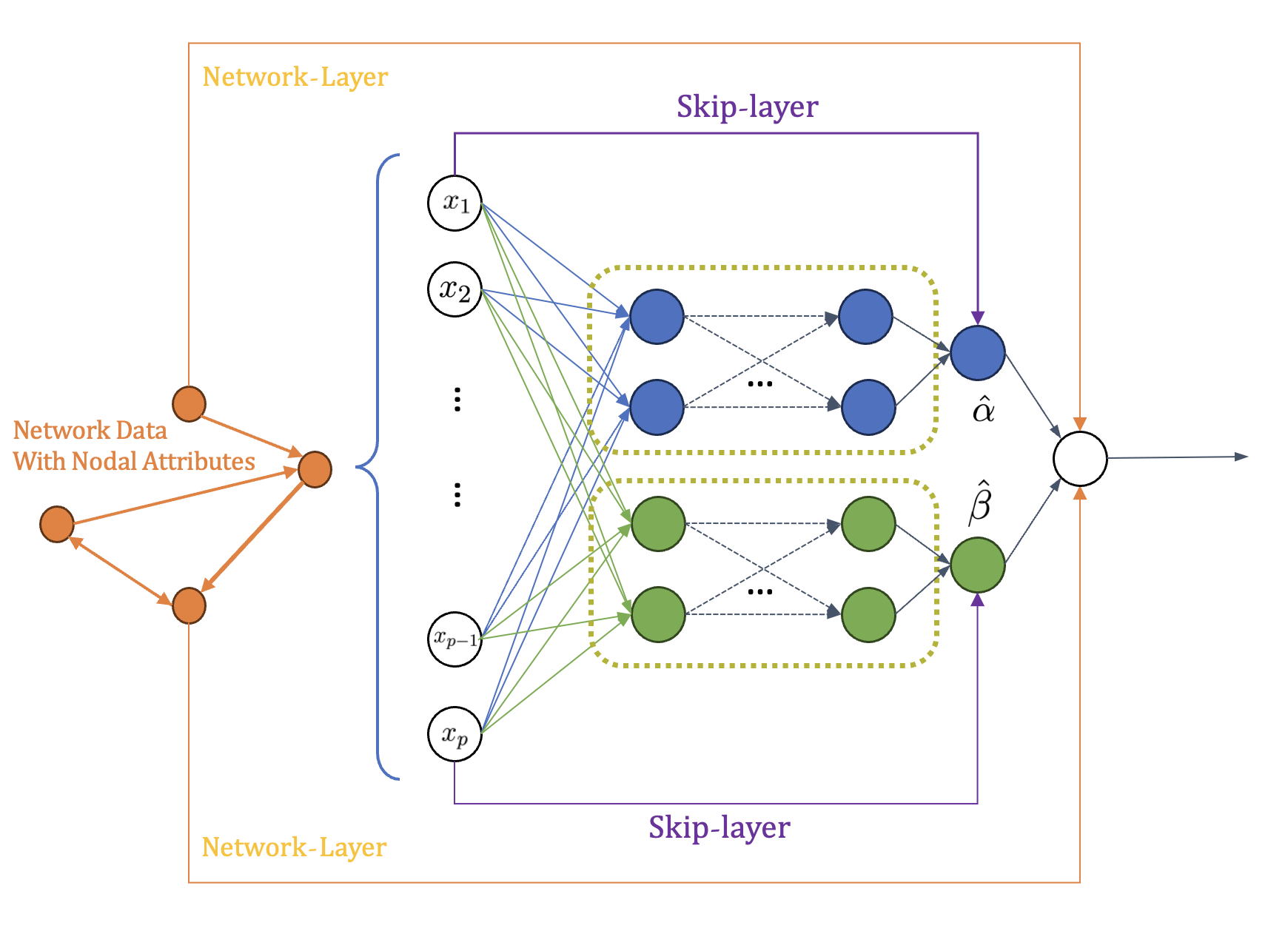}
    \caption{NetworkNet Methodology}
    \label{NetworkNetPlot}
\end{figure}

The architecture of NetworkNet is shown in Figure \ref{NetworkNetPlot}. The algorithm starts with the inputs of the count-valued networks with high-dimensional nodal attributes on the left-hand side. Each dimension of the nodal attributes is treated as the input of two separate parts of the DNNs in dash boxes, where the weights of blue and green linkages are trained within deep neural networks to approximate the latent expansiveness function $\alpha(\cdot)$ and popularity function $\beta(\cdot)$, respectively. Meanwhile, two skip layers are introduced to select nodal attributes in each iterating update in Algorithm \ref{Algorithm1}, where a hierarchical proximal operator described in Algorithm \ref{alg:hierarchical_prox} is used for specially designed constraints in \eqref{eq:optimization_problem}. As NetworkNet models nodal heterogeneity with random network models with high-dimensional nodal attributes, we have the additional Network Layer marked in yellow demonstrates that the observed count-valued network structures are included in the empirical loss function defined in \eqref{empiricalLoss}. 
As illustrated in Figure \ref{NetworkNetPlot}, the architecture of NetworkNet is specifically designed to unify advanced principles from statistical random network models and deep learning. This synergistic approach offers a robust methodology capable of explicitly modeling complex nodal heterogeneity within random networks characterized by high-dimensional attributes.

\subsection{Algorithms}

We propose an efficient and stable alternative optimization algorithm to effectively train the specially designed deep neural networks in Figure \ref{NetworkNetPlot} with the objective function in \eqref{eq:optimization_problem}.  The core idea is to utilize the hierarchical proximal operator to shrink the first-layer weights of neural networks, and decouple the approximations of expansiveness function $\alpha(\cdot)$ and popularity function $\beta(\cdot)$ into two iterative steps.

\begin{algorithm}[h]
\caption{NetworkNet Algorithm}
\label{Algorithm1}
\begin{algorithmic}[1]
\spacingset{1}
\REQUIRE Adjacency matrix $\bm{A} \in \N^{n \times n}$, nodal attributes matrix $\bm{X} \in \R^{n \times p}$; Hyper-parameters $\lambda_1, \lambda_2, \gamma, M,$ learning rate $\rho$, tolerance $\epsilon$
\STATE Initialization of  $(\hat{\bm{\alpha}}^{(0)}, \hat{\bm{\beta}}^{(0)})$ 
\FOR{$t = 0$ to $T_{\text{max}}-1$} 
    \STATE $(\hat{\bm{\beta}}^{(t+1)}, (\bm{W}_\beta, \bm{\theta}_\beta)) \gets \text{Update}(g, \bm{X}, \bm{A}, \hat{\bm{\alpha}}^{(t)}, \lambda_2, \gamma, \rho, M)$ for $\beta$
    \STATE $(\hat{\bm{\alpha}}^{(t+1)}, (\bm{W}_\alpha, \bm{\theta}_\alpha)) \gets \text{Update}(f, \bm{X}, \bm{A}, \hat{\bm{\beta}}^{(t+1)}, \lambda_1, \gamma, \rho, M)$ for $\alpha$
    \IF{$\|\hat{\bm{\alpha}}^{(t+1)} - \hat{\bm{\alpha}}^{(t)}\|_2  < \epsilon$ \textbf{and} $\|\hat{\bm{\beta}}^{(t+1)} - \hat{\bm{\beta}}^{(t)}\|_2 < \epsilon$}
        \STATE \textbf{break}
    \ENDIF
\ENDFOR
\STATE $\hat{\bm{\alpha}} \gets \hat{\bm{\alpha}}^{(t+1)}$, $\hat{\bm{\beta}} \gets \hat{\bm{\beta}}^{(t+1)}$
\STATE $\mathcal{S}_\alpha \gets \{ k : |(\bm{\theta}_\alpha)_k| > 0 \}$, $\mathcal{S}_\beta \gets \{ k: |(\bm{\theta}_\beta)_k| > 0 \}$
\RETURN Estimated parameters $\hat{\bm{\alpha}}, \hat{\bm{\beta}} $; Selected feature sets $\mathcal{S}_\alpha, \mathcal{S}_\beta$  
\end{algorithmic}
\end{algorithm}

Specifically, we model two types of nodal heterogeneity via two separate neural networks $f(\cdot; \bm{W}_\alpha, \bm{\theta}_\alpha)$ and $g(\cdot; \bm{W}_\beta, \bm{\theta}_\beta)$ from the class $\mathcal{H}$ with skip-layers. Let $\bm{\alpha} = [\alpha(\bm{x}_1), \dots, \alpha(\bm{x}_n)]^\top$ and $\bm{\beta} = [\beta(\bm{x}_1), \dots, \beta(\bm{x}_n)]^\top$ represent the true nodal heterogeneity parameter vectors, and denote the estimated nodal heterogeneity parameter vectors as $\hat{\bm{\alpha}}=[f(\bm{x}_1),\cdots,f(\bm{x}_n)]^\top$ and $\hat{\bm{\beta}}=[g(\bm{x}_1),\cdots,g(\bm{x}_n)]^\top$. The iterative optimization procedures are straightforward as demonstrated in Algorithm \ref{Algorithm1}: 
we update the network parameters $(\bm{\theta}_\beta,\bm{W}_\beta)$ for $g(\cdot)$ holding the $\hat{\bm{\alpha}}$ fixed; and then  update the network parameters $(\bm{\theta}_\alpha,\bm{W}_\alpha)$ for $f(\cdot)$ holding the $\hat{\bm{\beta}}$ fixed. The details of optimization in these iterative updates are given in Algorithm \ref{Algo2}. As the hierarchy constraints are separable over the features but the identifiability constraint is related to both two updates, we 
optimize the objective in \eqref{eq:optimization_problem} by constrained proximal gradient descent, which is given in Algorithm \ref{alg:hierarchical_prox}.

\begin{algorithm}[h]
\caption{The Update Procedure \footnotesize (Steps 3 \& 4 in Algorithm \ref{Algorithm1})}
\label{Algo2}
\begin{algorithmic}[1]
\spacingset{1}
\REQUIRE $f, g, \bm{X}, \bm{A}, \bm{\lambda}_{\text{fixed}}, \lambda, \gamma, \rho, M$, let $\bm{W}^{(1)} \in \R^{p \times d_1}$ be the first layer weights
\STATE Let $(\bm{W}, \bm{\theta})$ be the parameters of network $f$.
\IF{to update $\alpha$}
    \STATE Define loss function for $\alpha$:
    $$\mathcal{L}(\bm{W}, \bm{\theta}) = \sum_{i \neq j} \left[ e^{L_\alpha} - A_{ij}L_\alpha \right] + \lambda \|\bm{\theta}\|_1 + \gamma \left[\sum_k f_{\bm{W}}(\bm{x}_k) - \sum_k \lambda_{\text{fixed},k}\right]^2$$
    where $L_\alpha=f_{\bm{W}}(\bm{x}_i) + \lambda_{\text{fixed},j}$.
\ELSE
    \STATE Define loss function for $\beta$:
    $$\mathcal{L}(\bm{W}, \bm{\theta}) = \sum_{i \neq j} \left[ e^{L_\beta} - A_{ij}L_\beta \right] + \lambda \|\bm{\theta}\|_1 + \gamma \left[\sum_k \lambda_{\text{fixed},k} - \sum_k g_{\bm{W}}(\bm{x}_k)\right]^2$$
    where $L_\beta=\lambda_{\text{fixed},i} + g_{\bm{W}}(\bm{x}_j)$.
\ENDIF
\FOR{$k = 1$ to $T_{\text{max}}$}
    \STATE Gradient descent: $(\bm{W}, \bm{\theta}) \gets (\bm{W}, \bm{\theta}) - \rho \nabla \mathcal{L}(\bm{W}, \bm{\theta})$
    \STATE Proximal step: $(\bm{W}^{(1)}, \bm{\theta}) \gets \text{HierarchicalProx}(\bm{W}^{(1)}, \bm{\theta}, \rho\lambda, M)$
\ENDFOR
\STATE $\hat{\bm{\lambda}}_{\text{new}} \gets [f(\bm{x}_1), \dots, f(\bm{x}_n)]^\top$
\STATE \textbf{Return:} $\hat{\bm{\lambda}}_{\text{new}}, (\bm{W}, \bm{\theta})$
\end{algorithmic}
\end{algorithm}
 
\begin{algorithm}[]
\caption{Hierarchical Prox Procedure \footnotesize (Step 9 in Algorithm \ref{Algo2})}
\label{alg:hierarchical_prox}
\begin{algorithmic}[1]
\spacingset{1}
\REQUIRE 
    Parameters $\bm{W}$, $\bm{\theta}, \tau$, $M$      
\FOR{$k = 1$ to $p$} 
    \STATE $\theta_{\text{new}, k}  \gets \text{sign}(\theta_k) \cdot \max(|\theta_k| - \tau, 0) $  
    \STATE Let $\bm{w}_{k, \cdot}$ be the $k$-th row of $\bm{W}^{(1)}$, then update
$$\bm{w}_{k, \cdot, \text{new}} \gets \bm{w}_{k, \cdot}   \min \left(1, \frac{M \cdot |\theta_{\text{new}, k}|}{\| \bm{w}_{k, \cdot} \|_2   } \right)$$
\ENDFOR
\RETURN $(\bm{W}^{(1)}_{\text{new}}, \bm{\theta}_{\text{new}})$ 
\end{algorithmic}
\end{algorithm}

There are three hyper-parameters in the NetworkNet optimization: the regularization parameters $\lambda_1$ and $\lambda_2$, and the balance parameter $M$  that governs the trade-off between the neural-net components and skip layers.  We employ the High-dimensional Bayesian Information Criterion (HBIC,  \citealp{wang2013calibrating}) for hyperparameter tuning in practical applications with large-scale networks, where parallel computing is applicable.

\subsection{Theoretical Results}
\label{TheoResults}

The nodal heterogeneity functions $\alpha(\cdot)$ and $\beta(\cdot)$ of nodal attributes $\bm{X}$ can be complex and highly non-linear. Below, we consider them belonging to a general class of functions that are Hölder smooth and derive the non-asymptotic upper bound of the NetworkNet approximation. We prove that, with mild assumptions, the proposed NetworkNet can consistently approximate the latent nodal heterogeneity functions and give consistent estimations of individual expansiveness and popularity.

\begin{definition}[H\"{o}lder smooth function]
A function $f: \mathbb{R}^s \rightarrow \mathbb{R}$ is said to be $\gamma$-Hölder smooth if all $\lfloor \gamma \rfloor$-order partial derivatives of $f$ exist and satisfies the H\"{o}lder condition
$$\left |f^{(\lfloor \gamma \rfloor)}(x) - f^{(\lfloor \gamma \rfloor)}(y)\right| \leq a \left \|x - y \right\|_2^{\gamma - \lfloor \gamma \rfloor}$$
for some constant $a$.
\end{definition}

\begin{definition}[H\"{o}lder smooth compositional function]
\label{def:holder}
Consider the following collection of parameters with $J \in \mathbb{N}_+$: dimension vectors $\mathbf{k} = (k_1, \dots, k_{J+1})^\top \in \mathbb{N}_+^{J+1}$, intrinsic dimension vectors $\mathbf{s} = (s_1, \dots, s_J)^\top \in \mathbb{N}_+^{J}$, smoothness parameters $\boldsymbol{\gamma} = (\gamma_1, \dots, \gamma_J)^\top \in \mathbb{R}_+^{J}$, and domain bounds $\mathbf{a} = (a_1, \dots, a_J)^\top \in \mathbb{R}^{J}$ and $\mathbf{b} = (b_1, \dots, b_J)^\top \in \mathbb{R}^{J}$.  A function $f : [a_1, b_1]^{k_1} \to \mathbb{R}^{k_{J+1}}$ belongs to the class $\mathcal{F}$ of \emph{H\"{o}lder smooth compositional functions} if it admits the representation
\(
    f(\mathbf{x}) = (e_J \circ \cdots \circ e_2 \circ e_1)(\mathbf{x}), \quad \forall\, \mathbf{x} \in [a_1, b_1]^{k_1},
\)
where for each $u = 1, \dots, J$, the map $e_u : [a_u, b_u]^{k_u} \to [a_{u+1}, b_{u+1}]^{k_{u+1}}$ is constructed component-wise: each component $e_{uv} : [a_u, b_u]^{s_u} \to [a_{u+1}, b_{u+1}]$ is a $\gamma_u$-H\"{o}lder smooth function that depends on at most $s_u$ of the $k_u$ input coordinates.
\end{definition}

\begin{assumption}[Smoothness of nodal heterogeneity functions]
\label{smoothness}
Denote $\mathcal{F}$ as the class of $\gamma_u$-Hölder smooth compositional functions in terms of nodal attributes $x_u, \forall u \in [p]$. The true nodal heterogeneity function $\alpha_0,\beta_0 \in \mathcal{F}$. 
\end{assumption} 

Denote $\eta_0 = (\alpha_0,\beta_0)\in\mathcal{G}(\mathcal{F})$ as the true nodal heterogeneity function with both nodal expansiveness and nodal popularity, where $\mathcal{G}(\mathcal{F}):=\{(\alpha,\beta):\alpha \in \mathcal{F}, \beta\in\mathcal{F}\}$. Assumption \ref{smoothness} requires that the complex latent nodal heterogeneity functions are relatively smooth in terms of the nodal attributes. The mild assumption gives a wide class of functions to model the relationship between nodal attributes and nodal heterogeneity \citep{liu2020deep, kohler2021rate, schmidt2020nonparametric}. 

\begin{assumption}[Sparsity of nodal attributes]
\label{sparsityAsp}
    Assume the subset of true relevant nodal attributes satisfies $$|\mathcal{A_\alpha}|+|\mathcal{A_\beta}| = r_\alpha + r_\beta \ll p$$
    where $r_\alpha + r_\beta = o (\sqrt{n}\log^2(n) )$ and $p = O\left(n^3\log^3(n)\right)$. 
\end{assumption}  

For network data with rich nodal information, we usually consider that the high-dimensional true nodal attributes are sparse. As both the true relevant features and dimensions of nodal attributes can increase with sample size, and the fraction of true nodal attributes is $(r_\alpha + r_\beta)/p = o\left(n^{-5/2}\log^{-1}(n)\right)$, Assumption \ref{sparsityAsp} gives a mild requirement about the sparsity of nodal attributes. 

\begin{assumption}[DNN technical conditions]
\label{as2_DNN}
    There exists an $L$-layer ReLU-DNN class $\mathcal{H}$ with layer widths $\{d_v\}_{v=2}^L$ and minimum width $W = \min_{v=2,\cdots,L}d_v$ such that 
    \begin{enumerate}
        \item $L> J + 216( 
    \sum_{u=1}^J \lceil \gamma_u^2 \rceil+1);$
        \item $ W \geq 81 d_L \max_{1 \le u \le J} r_u (\lceil \gamma_u \rceil + s_u + 2)^{s_u+1} 3^{s_u+1};$
        \item $  L \max_{2 \le v \le L} d_v \asymp L \min_{2 \le v \le L} d_v \asymp n^{\frac{\tilde{s}}{2( 2 \tilde{\gamma} + \tilde{s})}}$, where $\tilde{\gamma} = \tilde{\gamma}_{\tilde{u}}$ and $\tilde{s} = s_{\tilde{u}}$, in which the index $\tilde{u}$ is defined by $\tilde{u} = \arg \min_{1 \le u \le J} \frac{\tilde{\gamma}_u}{s_u}$ with $\tilde{\gamma}_u = \gamma_u \prod_{v=u+1}^{J} (\gamma_v \land 1).$ 
    \end{enumerate}
\end{assumption} 

Assumption \ref{as2_DNN} is a commonly used technical assumption for DNNs \citep{schmidt2020nonparametric, liu2020deep}. With the width and length of neural networks increasing with sample size, NetworkNet can approximate the Hölder nodal heterogeneity functions based on nodal attribute matrix $\bm{X}\in\mathbb{R}^{n\times p}$ with the non-asymptotic bound given in Theorem \ref{ConsistencyEsti}.

\begin{theorem} 
\label{ConsistencyEsti}
Denote the estimated nodal heterogeneity functions as $\hat{\eta}=(\hat{\alpha},\hat{\beta})$ and the true nodal heterogeneity functions as  $\eta_0=( \alpha_0 , \beta_0 )$. Under Assumption \ref{smoothness}-\ref{as2_DNN} and \ref{ExtraAssumption} in the Supplements, one has
$$
\sup_{\eta_0 \in \mathcal{G}(\mathcal{F})} \|\hat{\eta}-\eta_0\|_{n,2} = O_p\left( n^{-\frac{\tilde{\gamma}}{2\tilde{\gamma} + \tilde{s}}}\log^2(n)\right)
$$
\end{theorem}

Theorem \ref{ConsistencyEsti} shows the consistency in estimation of nodal heterogeneity functions, where the convergence rate achieves the classical nonparametric regression rate \citep{schmidt2020nonparametric}. Based on consistent approximation of nodal heterogeneity functions, NetworkNet can give a consistent estimation for each node about its expansiveness and popularity. 
 
Furthermore, we explore the general approximation error bound of expansiveness and popularity function over the compact attribute space. Denote $\mathcal{A}=\mathcal{A}_\alpha\cup\mathcal{A}_\beta$ represent the active subspace, and let $r=r_\alpha+r_\beta$, then $|\mathcal{A}|\leq r \ll p$ based on Assumption \ref{sparsityAsp}. With the following mild assumptions, we provide a general non-asymptotic bound for nodal heterogeneity approximation in Theorem \ref{A:thm_general}. 

\begin{assumption}[Design regularity on active subspace]
\label{A:design_regularity}
The fixed design points $\{\bm{x}_1,\dots,\bm{x}_n\}$ satisfy the following with respect to Borel reference measure $\mu$ and active subspace $\mathcal{A}$:
\begin{enumerate}[label=(\roman*)]
    \item Let $\mathcal{X}_{\mathcal{A}}=\{\bm{x}_{\mathcal{A}}:\bm{x}\in\mathcal{X}\}$ be the projection onto the active subspace. The fill distance 
    $h_{n,\mathcal{A}} = \sup_{\bm{z}\in\mathcal{X}_{\mathcal{A}}}\min_{1\leq i\leq n}\|\bm{z}-(\bm{x}_i)_{\mathcal{A}}\|_2$
    satisfies $h_{n,\mathcal{A}} = O(n^{-1/r})$. 
    
    \item There exists $C_\mu>0$ (independent of $n$) such that the Voronoi cells 
    $$V_i = \bigl\{\bm{x}\in\mathcal{X}: \|(\bm{x})_{\mathcal{A}}-(\bm{x}_i)_{\mathcal{A}}\|_2 \leq \|(\bm{x})_{\mathcal{A}}-(\bm{x}_j)_{\mathcal{A}}\|_2,\;\forall j\neq i\bigr\}$$
    satisfy $\max_{1\leq i\leq n}\mu(V_i)\leq C_\mu/n$.
\end{enumerate}
\end{assumption}

\begin{assumption}[Lipschitz regularity of the neural network class]
\label{A:lipschitz_DNN}
There exists a sequence $b_n>0$ such that every $f\in\mathcal{H}$ satisfying the constraints in optimization problem~(4) is $b_n$-Lipschitz on $\mathcal{X}_{\mathcal{A}}$:
$$
|f(\bm{x})-f(\bm{x}')| \leq b_n\|(\bm{x})_{\mathcal{A}}-(\bm{x}')_{\mathcal{A}}\|_2, \quad \forall\,\bm{x},\bm{x}'\in\mathcal{X},
$$
with $b_n = O\bigl(n^{\rho}/\log^2(n)\bigr)$, where $\rho = \tilde{\gamma}/(2\tilde{\gamma}+\tilde{s})$.
\end{assumption}
 
It is worth noting that Assumption~\ref{A:design_regularity} is related to unknown active set $\mathcal{A}$ only, where the rate $n^{-1/r}$ is mild and commonly-used \citep{Wendland_2004, narcowich2005sobolev, pronzato2023quasi} for $s$-dimensional space.  In practice, it requires that the nodal attributes are reasonably spread over the low-dimensional active subspace, while the nodal attributes neither cluster nor leave large gaps.

\begin{theorem} 
\label{A:thm_general}
Denote the estimated nodal heterogeneity functions as $\hat{\eta}=(\hat{\alpha},\hat{\beta})$ and the true functions as $\eta_0=(\alpha_0,\beta_0)$. Under Assumptions in Theorem \ref{ConsistencyEsti} and the additional Assumptions \ref{A:design_regularity} and \ref{A:lipschitz_DNN}, one has   
\begin{equation}
\sup_{\eta_0\in\mathcal{G}(\mathcal{F})} \|\hat{\eta}-\eta_0\|_{L_2} = O_p\!\left(n^{-\frac{\tilde{\gamma}}{2\tilde{\gamma} + \tilde{s}}}\log^2(n) + \frac{n^{\tau}}{\log^2(n)}\right),
\label{A:explicit_rate}
\end{equation}
where $\tau = \tilde{\gamma}/(2\tilde{\gamma}+\tilde{s}) - 1/r $.
\end{theorem}

We explain the conclusion in Theorem \ref{A:thm_general} in several regimes in the following Corollary \ref{A:corollary_regimes}.
\begin{corollary}
\label{A:corollary_regimes}
Let $\rho = \tilde{\gamma}/(2\tilde{\gamma}+\tilde{s})$. Under the assumptions of Theorem~\ref{A:thm_general}, the following regimes hold.
\begin{enumerate}[label=(\alph*)]
    \item If $\tau < -\rho$,  or equivalently $2\tilde{\gamma}\,r < 2\tilde{\gamma}+\tilde{s},$ then $n^{\tau}/\log^2(n) = o(n^{-\rho}\log^2(n))$. Theorem \ref{A:thm_general} and Theorem \ref{ConsistencyEsti} share the same rate. 
    \item If $-\rho < \tau < 0$,  or equivalently $2\tilde{\gamma}+\tilde{s} < 2\tilde{\gamma}\,r < 2(2\tilde{\gamma}+\tilde{s})$, then
    $$
    \|\hat{\eta}-\eta_0\|_{L_2} = O_p\!\left(\frac{n^{\tau}}{\log^2(n)}\right) = O_p\!\left(\frac{n^{\rho-1/r}}{\log^2(n)}\right).
    $$ 
\end{enumerate}
\end{corollary}

\begin{remark}
    The condition $2\tilde{\gamma}\,r < 2\tilde{\gamma}+\tilde{s}$, equivalently $r < 1+\tilde{s}/(2\tilde{\gamma})$, is a joint condition on the smoothness parameters $(\tilde{\gamma},\tilde{s})$ and the total number of active features $r$. For a fixed $r$, the condition $2\tilde{\gamma}\,r < 2\tilde{\gamma}+\tilde{s}$ requires $\tilde{s}$ to be large relative to $\tilde{\gamma}$. 
\end{remark}

\section{Simulations}
To evaluate the numerical performance of NetworkNet on both estimation of nodal heterogeneity and nodal attribute selection, we generate the random count-valued directed network from model \eqref{ModelSetting} with size $n=100$. The nodal attribute matrix $\bm{X} \in \mathbb{R}^{n \times p}$ is then generated independently from a uniform distribution, $x_{ij} \sim \text{Unif}(-1, 1), i\in[n], j\in[p]$ with dimension $p=1000$. As shown in \eqref{eq:model_main}, the complex nodal heterogeneity, including both expansiveness and popularity, is determined by high-dimensional nodal attributes. To fully show the effectiveness of NetworkNet in modeling complex nodal heterogeneity, we consider the following two different settings to simulate the expansiveness function $\alpha(\cdot)$ and popularity function $\beta(\cdot)$: a linear setting
        \begin{equation}
        \alpha(\bm{x}_i) =  \sum_{k=1}^{5} x_{i,k}, \text{    } \beta(\bm{x}_j) =  \sum_{k=6}^{10} x_{j, k} \label{LinaerSim}  
        \end{equation}  
and a nonlinear setting
        \begin{align}
        \alpha(\bm{x}_i) = 5 \left[ |x_{i,1}| + |x_{i,2}| + \log(|x_{i,3}|) + \log(|x_{i,4}| + |x_{i,5}| )\right] \nonumber  \nonumber \\
        \beta(\bm{x}_j) = 5 \left[ |x_{j,6}| + |x_{j,7}| + \log(|x_{j,8}|)  +  \log(|x_{j,9}|  + |x_{j,10}|) \right] \label{nonLinaerSim}
        \end{align}   
As shown in equations \eqref{LinaerSim} and \eqref{nonLinaerSim}, we set the true nodal attributes for $\alpha$ as $x_1$ to $x_5$, and true nodal attributes for $\beta$ are from $x_6$ to $x_{10}$. 
Correspondingly, the indicator of the true subset of $\alpha$-related nodal attributes is $\mathcal{A}_\alpha=\{1,2,3,4,5\}$, and indicator of the true subset of $\beta$-related nodal attributes is $\mathcal{A}_\beta=\{6,7,8,9,10\}$ in all cases. 

We generate $r = 1,2,\cdots,100$ replications of each of the simulation settings above. Denote the estimated nodal \textit{expansiveness} and \textit{popularity} for node $i\in[n]$ in $r$-th round of replications as $\hat{\alpha}_i^{(r)}$ and $\hat{\beta}_i^{(r)}$, respectively. Denote the estimated subsets of $\alpha$-related features and $\beta$-related features as $\hat{\mathcal{A}}_\alpha^{(r)}$ and $\hat{\mathcal{A}}_\beta^{(r)}$ respectively. With known true nodal heterogeneity $(\bm{\alpha}_0, \bm{\beta}_0)$ and true subset of nodal attributes $\mathcal{A}_\alpha, \mathcal{A}_\beta$, we evaluate and compare different approaches with the following classic evaluations:  To assess the performance in estimating the nodal heterogeneity for each method, we compute the average root mean squared error for both $\alpha$ and $\beta$ for comparison based on the simulation results; to evaluate the performance in selecting relevant nodal attributes, we consider the precision, true-positive-rate and F1 value for both expansiveness-related selection and popularity-related selection. The definitions of these evaluations are shown in Table \ref{DefMeas}. 

\begin{table}[htp!]
\spacingset{1}
\centering
\caption{  The definitions of the simulation performance measures to assess the performance of estimation of $\alpha(\cdot)$ and $\beta(\cdot)$, and selection of nodal attributes.} 
\label{DefMeas}
\footnotesize 
\begin{tabular}{@{}lll@{}}
\toprule
\textbf{Measures}             & \textbf{Notation} & \textbf{Formula} \\ \midrule
\ul{ \textbf{Estimation}}     &                   &                  \\
Average root mean squared error for $\alpha(\cdot)$               & $\alpha$-RMSE               &   $\{100^{-1} \sum_{r} [\frac{1}{n} \sum_{i=1}^n(\hat{\alpha}_i^{(r)}-\alpha_{i,0})^2]\}^{1/2}$      \\
Average root mean squared error for $\beta(\cdot)$          & $\beta$-RMSE               & $\{100^{-1} \sum_{r}  [\frac{1}{n} \sum_{i=1}^n(\hat{\beta}_i^{(r)}-\beta_{i,0})^2]\}^{1/2}$                \\
\ul{ \textbf{Selection}}      &                   &                  \\
Average precision for $\alpha$-related  attributes           & $\alpha$-Precision                &   $100^{-1} \sum_{r} |\hat{\mathcal{A}}_\alpha^{(r)} \cap \mathcal{A}_\alpha|/|
\hat{\mathcal{A}}_\alpha^{(r)}|$               \\
Average precision for $\beta$-related  attributes  & $\beta$-Precision                & $100^{-1} \sum_{r} |\hat{\mathcal{A}}_\beta^{(r)} \cap \mathcal{A}_\beta|/|
\hat{\mathcal{A}}^{(r)}_\beta|$        \\
Average true positive rate of $\alpha$-related attributes &         $\alpha$-TPR      &     $100^{-1} \sum_{r}  |\hat{\mathcal{A}}^{(r)}_\alpha \cap \mathcal{A}_\alpha |/|
\mathcal{A}_\alpha |$      \\ 
Average true positive rate of $\beta$-related  attributes &  $\beta$-TPR               &   $100^{-1} \sum_{r}  |\hat{\mathcal{A}}^{(r)}_\beta \cap \mathcal{A}_\beta |/|
\mathcal{A}_\beta |$               \\
Average F1 value for $\alpha$-related attributes &         $\alpha$-F1      &     $100^{-1} \sum_{r=1}^{100}   2 \cdot |\hat{\mathcal{A}}_\alpha^{(r)} \cap \mathcal{A}_\alpha| / (|\hat{\mathcal{A}}_\alpha^{(r)}| + |\mathcal{A}_\alpha|)  $       \\ 
Average F1 value for $\beta$-related attributes &  $\beta$-F1               &   $100^{-1} \sum_{r=1}^{100}  2 \cdot |\hat{\mathcal{A}}_\beta^{(r)} \cap \mathcal{A}_\beta| / (|\hat{\mathcal{A}}_\beta^{(r)}| + |\mathcal{A}_\beta|)$            \\
\bottomrule
\end{tabular}
\begin{flushleft}
\footnotesize 
Note: The parameter $r$ refers to the $r$-th replication of the simulations.
\end{flushleft} 
\end{table}

As there are no specialized approaches in previous literature to model the nodal heterogeneity in a count-valued random networks with high-dimensional nodal attributes that can also select the nodal attributes simultaneously, we consider the following 2 two-stage combinations of classic statistical and deep learning algorithms as competitors: (1) The classic Lasso estimator \citep{Tibshirani1996} based on maximum likelihood estimates (denoted as ``MLE+Lasso'') and (2) The DNN estimator Lassonet \citep{lemhadri2021lassonet} based on maximum likelihood estimates (denoted as ``MLE+LassoNet''). 
For both of the two-stage methods, the maximum likelihood estimates for the two parameters of each individual node are numerically approximated based on model \eqref{ModelSetting} with observed network structures. Then the estimated vectors of nodal expansiveness and popularity serve as the responses in both Lasso and Lassonet algorithms for nodal attribute selection. For all simulation cases, the network architectures are set to be $(32,16)$ and training epochs are set to be $1000$ with random initialization. The tuning parameters are chosen by high-dimensional Bayesian information criteria, and the optimal models are used to approximate the nodal heterogeneity function and estimate the nodal heterogeneity $\hat{\bm{\alpha}}$ and $\hat{\bm{\beta}}$.

Tables \ref{LinearSim} and \ref{MomLinearSim} summarize the simulation results under linear and non-linear settings, respectively, and demonstrate the superior performance of the proposed method, NetworkNet, in both nodal heterogeneity estimation and nodal attribute selection. 

For the estimation of nodal heterogeneity, we use RMSE to quantify the distance between estimated nodal heterogeneity parameters $(\hat{\bm{\alpha}}, \hat{\bm{\beta}})$ and true expansiveness $\bm{\alpha}$ and popularity $\bm{\beta}$. In all simulation settings, the maximum likelihood estimator has the largest average RMSE as it uses only information about network structure. Meanwhile, all the other three approaches benefit from the additional nodal attribute information. Specifically, the LassoNet method, which includes Lasso as a special case, gives a smaller RMSE compared to Lasso-based maximum likelihood estimations when the latent nodal heterogeneity is non-linear. The proposed NetworkNet gives the estimations of heterogeneity parameters with much lower RMSE in both linear and nonlinear nodal heterogeneity settings. 

For the selection of nodal attributes, we evaluate the performance by comparing to the known true subset. 
As the maximum likelihood estimation does not perform selection, all the evaluations of attribute selection for MLE are blank (/). 
As the simulation results show, the proposed NetworkNet can select the nodal attributes accurately with precision, TPR, and F1 score close to 1. 
While both the two-stage methods, MLE+Lasso and MLE+Lassonet, cannot select the true nodal attributes, they usually have good performance in classic independent data. One of the reasons is that both two-stage estimates are highly dependent on the first stage MLE, and cannot capture the complex nodal heterogeneity based on nodal attributes in practice. However, the proposed NetworkNet, combining the statistical model of random networks and deep neural network techniques, can effectively model the nodal heterogeneity while accurately selecting nodal attributes. 

\begin{table}[ht]
\spacingset{1}
\caption{Performance of estimation and nodal attribute selection for linear nodal heterogeneity}
\label{LinearSim}
\begin{tabular}{@{}cccccc@{}}
\toprule
\textbf{}                          &                    & \textbf{NetworkNet} & \textbf{MLE+Lasso} & \textbf{MLE+LassoNet} & \textbf{MLE}    \\ \midrule
\multirow{4}{*}{\textbf{$\alpha$}} & \textbf{RMSE}      & 0.5870 (0.2623)     & 1.3196 (0.0835)    & 1.4785 (0.0941)       & 1.3758 (0.0833) \\
                                   & \textbf{Precision} & 0.9983 (0.0167)     & 0.0050 (0.0286)    & 0.0298 (0.0602)       & /               \\
                                   & \textbf{TPR}       & 0.9680 (0.1497)     & 0.0060 (0.0343)    & 0.0480 (0.0990)       & /               \\
                                   & \textbf{F1}        & 0.9737 (0.1218)     & 0.0055 (0.0312)    & 0.0365 (0.0739)       & /               \\
                                   \midrule
\multirow{4}{*}{\textbf{$\beta$}}  & \textbf{RMSE}      & 0.6163 (0.1142)     & 1.3081 (0.0876)    & 1.2672 (0.0896)       & 1.3654 (0.0885) \\
                                   & \textbf{Precision} & 0.7179 (0.1783)     & 0.0033 (0.0235)    & 0.0121 (0.0563)       & /               \\
                                   & \textbf{TPR}       & 1.0000 (0.0000)     & 0.0040 (0.0281)    & 0.0100 (0.0438)       & /               \\
                                   & \textbf{F1}        & 0.8230 (0.1252)     & 0.0036 (0.0256)    & 0.0106 (0.0471)       & /               \\
                                   \bottomrule
\end{tabular}

\vspace{1ex}
\footnotesize 
Note: The evaluations of attribute selection for MLE are blank (/) as the MLE method does not perform selection. 
\end{table}

\begin{table}[ht]
\spacingset{1}
\caption{Performance of estimation and nodal attribute selection for non-linear nodal heterogeneity}
\label{MomLinearSim}
\begin{tabular}{@{}cccccc@{}}
\toprule
\textbf{}                          &                    & \textbf{NetworkNet} & \textbf{MLE+Lasso} & \textbf{MLE+LassoNet} & \textbf{MLE}    \\ \midrule
\multirow{4}{*}{\textbf{$\alpha$}} & \textbf{RMSE}      & 1.8608 (0.3217)     & 4.0095 (0.3761)    & 3.6234 (0.4250)       & 5.2026 (0.4876) \\
                                   & \textbf{Precision} & 0.9105 (0.1791)     & 0.0131 (0.0468)    & 0.0503 (0.0646)       & /               \\
                                   & \textbf{TPR}       & 0.9880 (0.0844)     & 0.0180 (0.0642)    & 0.1620 (0.1943)       & /               \\
                                   & \textbf{F1}        & 0.9327 (0.1414)     & 0.0151 (0.0539)    & 0.0737 (0.0870)       & /               \\
                                   \midrule
\multirow{4}{*}{\textbf{$\beta$}}  & \textbf{RMSE}      & 2.0015 (0.4917)     & 4.2462 (0.4712)    & 3.7348 (0.3889)       & 5.5397 (0.7502) \\
                                   & \textbf{Precision} & 0.8291 (0.2872)     & 0.0150 (0.0480)    & 0.0375 (0.0760)       & /               \\
                                   & \textbf{TPR}       & 0.9300 (0.2564)     & 0.0220 (0.0690)    & 0.0540 (0.1058)       & /               \\
                                   & \textbf{F1}        & 0.8662 (0.2651)     & 0.0177 (0.0562)    & 0.0426 (0.0832)       & /               \\
                                   \bottomrule
\end{tabular}

\vspace{1ex}
\footnotesize 
Note: The evaluations of attribute selection for MLE are blank (/) as the MLE method does not perform selection. 
\end{table}

For nonlinear cases, the complex nodal heterogeneity enlarges the advantages of the proposed NetworkNet. Specifically, the MLE estimate, which is based only on network structure information, is less accurate compared to linear cases, which makes the MLE-based two-stage methods perform poorly on both estimation and nodal attribute selection. Meanwhile, NetworkNet approximates the nodal heterogeneity with the random network model and deep neural networks, which enables the capture of true nonlinear relationships and the accurate selection of the true subset of nodal attributes. As shown in the results in Table \ref{MomLinearSim}, NetworkNet gives a dominant performance in both nodal heterogeneity estimation and nodal attribute selection for non-linear settings.

\section{Empirical Analysis}
\label{EAsec}

Academic networks have been a longstanding focus in network science. In the statistics community, the increasing availability of collaboration and citation networks among statisticians \citep{JinJiashun2016AOAS,ji2022co} has spurred increasing interest in network centrality, community structures, research patterns, and disciplinary trends in the field \citep{gao2025JMLRaccepted,hayes2025co}. In this section, we study a related but distinct empirical question, emphasizing the dynamic evolution of influential research fields and their shifting prominence over time. 

With the powerful specialized framework of NetworkNet, we are equipped to more precisely quantify the scholars' nodal heterogeneity, using count-valued author-citation networks and incorporating individuals’ research interests as high-dimensional nodal attributes. Our empirical analysis reveals the influential research fields, selected by NetworkNet, that have largely affected the nodal heterogeneity through the expansiveness function and popularity function, respectively. Such nodal heterogeneity subsequently informs the latent interaction intensity that governs the generative mechanism of count-valued citation edges in author–citation networks.

\subsection{Dataset and Pre-processing}
The recently released dataset of large-scale academic network \citep{gao2023large}
consists of information about 97,436 publications authored by 168,171 scholars over a four-decade span in representative statistical journals.
In addition to the citation relationships, it records rich nodal information for each scholar. 
The availability of these nodal attributes enables deep exploration of scholarly interactions through the random networks. Furthermore, by selecting the influential nodal attributes within each decade, the dataset facilitates the analysis of the evolution and scholarly impact of various research topics over time. 

\begin{figure}[h]
    \centering
    \includegraphics[width=0.6\linewidth]{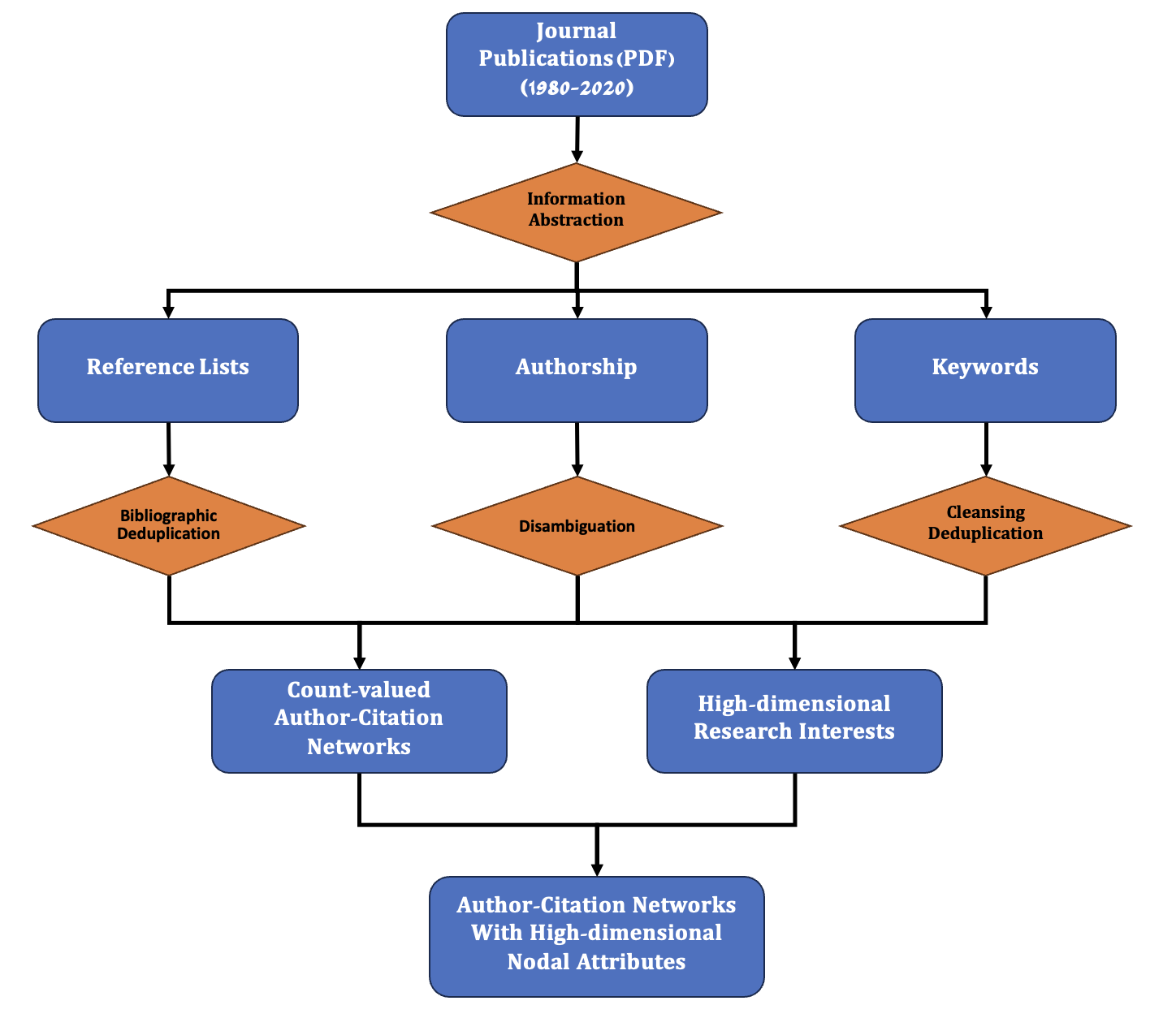}
    \caption{The construction of the academic author-citation networks}
    \label{DP}
\end{figure}

We first pre-process the raw data extracted from publications and construct the author-citation networks. As shown in Figure \ref{DP}, we start from the journal publications data where three parts of information are collected: authorship, reference lists, and also keywords. After the data cleaning, we construct the academic author-citation networks by combining the reference lists and the authorship, where the vertices represent different scholars and the weighted edges between them indicate the aggregated citation counts from one scholar to another. As the citation count increases once a new paper is published and the corresponding reference list is available, the edges in author-citation networks are naturally count-valued with directions from the author to the cited scholar. Meanwhile, for any scholars in the author-citation networks, we aggregate the keywords from all their publications within a time period as high-dimensional nodal attributes, which directly represent the distributions of both their research interests and contributions. 

\begin{figure}[]
    \centering
    \includegraphics[width=0.9\linewidth]{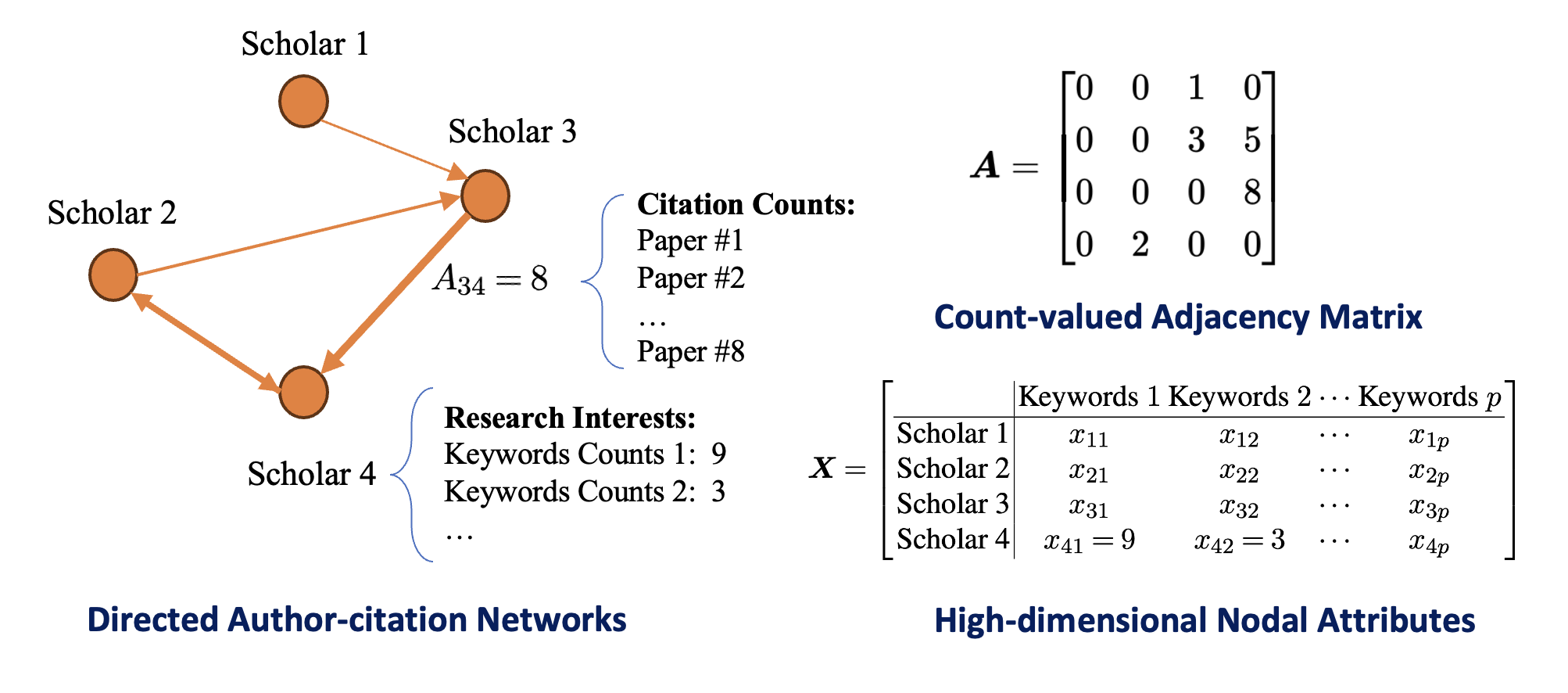}
    \caption{Example of count-valued author-citation networks with nodal attributes}
    \label{CollNetPlot}
\end{figure}

With the pre-processing and construction steps shown in Figure \ref{DP} and by splitting the publication-based raw data between 1981 and 2020 by decade, we construct four decade-specific author-citation networks with count-valued adjacency matrices and high-dimensional nodal attribute matrices. In the subsequent descriptions, we denote them as `80-90', `90-00', `00-10', and `10-20', respectively. 
Figure \ref{CollNetPlot} illustrates the specific data type associated with each decade, and the basic information about the four author-citation networks is summarized in Table \ref{NI}.
For each of the author-citation networks with $n$ scholars, we use adjacency matrix $\bm{A}\in\mathbb{N}^{n\times n}$ with non-negative integers to summarize the observed network structure, where element $A_{ij}$ represents the aggregated citation counts of scholar $j$ contributed by the publications written by author $i$. 
Nodal research profiles are encoded through a nodal attribute matrix $\bm{X}\in\mathbb{R}^{n\times p}$, where $p$ is the number of unique research keywords observed across all scholars, and $x_{ik}$ represents the appearance count of $k$-th keyword in author $i$'s publications.
Overall, we observe a rapid growth in the number of scholars and research keywords across the four decades. This expansion signifies the vibrant growth of the statistical discipline, accompanied by a progressive broadening and refinement of its core research areas.

\begin{table}[h]
\spacingset{1}
\centering
\caption{Network information of the four author-citation networks}
\label{NI}
\begin{tabular}{@{}lrrrr@{}}
\toprule
\multicolumn{1}{c}{\textbf{}}    & \multicolumn{1}{r}{\textbf{80-90}} & \multicolumn{1}{r}{\textbf{90-00}} & \multicolumn{1}{r}{\textbf{00-10}} & \multicolumn{1}{r}{\textbf{10-20}} \\ \midrule
\textbf{Publication counts}             & 11279                             & 18212                             & 25495                             & 38432                             \\
\textbf{Author counts}          
& 2967                              & 4998                              & 8283                              & 14347                             \\
\textbf{Keyword counts}            
& 393                               & 23761                             & 36400                             & 57927                             \\
\textbf{Edges counts}    & 78640                             & 211668                            & 447434                            & 632485                            \\
\textbf{Network density}        & 0.009                             & 0.008                             & 0.007                             & 0.003                             \\
\textbf{Average path length}       & 3.290                             & 3.113                             & 3.106                             & 3.336                             \\
\textbf{Diameter}                & 12                            & 13                            & 14                            & 13                           \\
\textbf{Clustering coefficient} & 0.185                             & 0.174                             & 0.152                             & 0.116                             \\
\textbf{Average degree}         & 26.505                            & 42.351                            & 54.018                            & 44.085                            \\ \bottomrule
\end{tabular}
\end{table}

\subsection{Empirical Findings} 
The directed edge in author-citation networks indicates the citation relationship. Within this framework, the \textit{expansiveness function} $\alpha(\cdot)$ describes the inherent characteristics of a scholar to cite the publications by diverse scholars and thereby generating outgoing citations broadly across the community. 
Conversely, the \textit{popularity function} $\beta(\cdot)$ captures a scholar's inherent attractiveness to receive more citations, which is also widely accepted as a measure of academic contributions. 

To analyze the evolution of influential research fields 
in statistics, we use the proposed NetworkNet to model the nodal heterogeneity in four decade-specific author-citation networks, respectively.  
We select 20 influential nodal attributes for \textit{expansiveness function} and \textit{popularity function}, respectively, by tuning the regularization parameters to control model sparsity.  
After fitting the NetworkNet models,
we evaluate the contributions of selected nodal attributes by their SHAP values \citep{lundberg2017unified}, a model-agnostic nodal attribute importance metric widely adopted in the machine learning community \citep{arenas2023complexityJMLR, wang2024feature}. 
Since the magnitudes of SHAP values are not comparable across the four author-citation networks, within-network SHAP rankings are used instead. 
Specifically, attributes are ranked by their SHAP values in each decade, and temporal trends in these ranks are analyzed for major statistical research fields across the four decades. 
The resulting ranks of all selected nodal attributes are reported in Table \ref{expansiveness-related} for nodal expansiveness and Table \ref{Beta-FeaturesTable} for nodal popularity, respectively. The decade-specific SHAP values of influential research fields are shown in Table \ref{T8090}-\ref{T1020} of the supplementary material.

As the selected top 20 keywords are all influential (top 0.1\% for the latter three decades), it is worth noting that the repeated selection across the four decades indicates the sustained importance of specific research fields. Table \ref{Allappear} summarizes the consistently selected influential research fields across all four decades, along with their rankings. As shown, \textit{bootstrap} and \textit{consistency} emerge as long-standing themes of substantial theoretical and practical scope, characterized by considerable depth and breadth. \textit{Bootstrap} methods offer great potential for general problems and often provide a tractable way for sufficiently general applications \citep{Diciccio1988Review, Horowitz2019Bootstrap}. Likewise, \textit{consistency}, as a foundational concept in statistics, always serves as a fundamental part of theoretical guarantees \citep{lin2014past}.  
Contributions to these fields have demonstrably influenced both the expansiveness and popularity of scholarly work over time.  
Furthermore, \textit{maximum likelihood} remains a method of enduring academic relevance due to its foundational role in statistical modeling and its widespread use in machine learning \citep{Kirch2025Challenges}
This is also evidenced by its consistent presence among the top 20 popularity-related nodal attributes across all four periods under study. This finding indicates that research contributions specifically focusing on maximum likelihood estimation significantly impact a scholar's citation attractiveness within the academic author-citation network. 

\begin{table}[h]
\spacingset{1}
\centering
\small
\caption{The ranks of the consistently selected influential research fields in all four decades}
\label{Allappear}
\begin{tabular}{@{}lrrrr@{}}
\toprule
\multicolumn{1}{c}{\textbf{Attributes}} & \multicolumn{1}{c}{\textbf{80-90}} & \multicolumn{1}{c}{\textbf{90-00}} & \multicolumn{1}{c}{\textbf{00-10}} & \multicolumn{1}{c}{\textbf{10-20}} \\ \midrule
\ul{ \textbf{Expansiveness-related Attributes} }                   & \multicolumn{4}{l}{\ul{}}                                                                                                                         \\
bootstrap                             & 2                                  & 1                                  & 2                                  & 6                                  \\
consistency                           & 7                                  & 7                                  & 7                                  & 10                                 \\
\ul{ \textbf{Popularity-related Attributes} }                   & \multicolumn{4}{r}{}                                                                                                                              \\
bootstrap                             & 2                                  & 1                                  & 2                                  & 4                                  \\
consistency                           & 6                                  & 6                                  & 8                                  & 15                                 \\
maximum likelihood estimation         & 16                                 & 15                                 & 19                                 & 10                                 \\
maximum likelihood                    & 17                                 & 3                                  & 10                                 & 19                                 \\ \bottomrule
\end{tabular}

\vspace{1ex}
\footnotesize 
Note: `80-90', `90-00', `00-10' and `10-20' represent the four decades \\ ``1981-1990'', ``1991-2000'', ``2001-2010'' and ``2011-2020'', respectively.
\end{table}

Table \ref{expansiveness-related} shows the rank changes of selected expansiveness-related nodal attributes over the four decades. Since expansiveness characterizes a node's inherent tendency to connect with other nodes in a network, in the author-citation networks, it represents a scholar's proclivity to generate outgoing citations broadly across the community. Scholars with higher expansiveness tend to cite publications by a wide and diverse set of scholars, producing more outward edges and larger out-degrees. Meanwhile, research fields with upward trends in the rank indicate a growing influence on scholarly expansiveness that reflects broader citation behavior.
This suggests that these fields are progressively evolving into broadly applicable methodological toolkits with strong extensibility and applicability.
For instance, the field of \textit{Bayesian inference} developed rapidly after the success of tackling the computing challenges with MCMC around 1990 \citep{gelfand1990sampling, brooks2011handbook}, and has become one of the primary areas in statistics \citep{ji2022co}. This trajectory aligns with the results in Table \ref{expansiveness-related} that  \textit{Bayesian inference} was not selected among influential attributes during the 1981–1990 period, but its relative position rose largely after 1990 from rank 18th to 11th. The increasing trend in its rank reflects the deepened research into Bayesian methods and their increased interdisciplinary integration with various fields.

Conversely, research fields exhibiting a declining rank over time indicate a reduced role in shaping nodal expansiveness, implying that scholars in these areas increasingly cite work from a narrower or more specialized author set.
This trend suggests that such areas are reaching methodological maturity and have formed a relatively stable scholarly community. 
For example, \textit{likelihood ratio test}, a long-standing statistical method, was a leading contributor to expansive citation behavior in earlier decades \citep{moreira2003conditional, arnold2024likelihood}, supported by its frequent use across diverse models and its strong theoretical properties \citep{kent1982robust, hogg1977probability}. 
Consistent with this, empirical results in Table \ref{expansiveness-related} show a strong expansiveness effect in the 1981–2000 interval. 
In later years, as its core theory became standardized and incorporated into textbooks, its relative contribution to generating broad citation links declined, becoming non-selected after 2000. Similar downward expansiveness-related rankings are observed for other mature methodologies, such as \textit{logistic regression}.

\begin{table}[htbp]\spacingset{0.9}
\centering
\small
\caption{The ranks of the selected influential research fields for nodal expansiveness in the four decades}
\label{expansiveness-related}
\begin{tabular}{@{}lllll@{}}
\toprule
\textbf{\textbf{Nodal expansiveness($\alpha$)-related fields}}                                  & \textbf{80-90}                     & \textbf{90-00}                      & \textbf{00-10}                        & \textbf{10-20}                      \\ \midrule   
simulation                     & 1    & 20   &    &      \\
bootstrap                      & 2    & 1    & 2  & 6    \\
reliability                    & 3    &      &    &      \\
logistic regression            & 4    & 8    & 15 &      \\
mean squared error             & 5    &      &    &      \\
empirical bayes                & 6    &      &    &      \\
consistency                    & 7    & 7    & 7  & 10   \\
poisson regression             & 8    &      &    &      \\
numerical integration          & 9    &      &    &      \\
successive difference analysis & 10   &      &    &      \\
maximum likelihood estimation  & 11   &      & 20 & 8    \\
repeated measures              & 12   &      &    &      \\
analysis of variance           & 13   &      &    &      \\
likelihood ratio test          & 14   & 15   &    &      \\
regression                     & 15   &      &    &      \\
kurtosis                       & 16   &      &    &      \\
nonparametric regression       & 17   & 2    & 3  &      \\
significance level             & 18   &      &    &      \\
2-way layout                   & 19   &      &    &      \\
hierarchical models            & 20   &      &    &      \\
maximum likelihood             &      & 3    & 12 & 19   \\
asymptotic normality           &      & 4    & 4  & 3    \\
robustness                     &      & 5    & 6  & 15   \\
EM algorithm                   &      & 6    & 5  & 2    \\
survival analysis              &      & 9    & 13 & 13   \\
missing data                   &      & 10   & 11 & 20   \\
Gibbs sampling                 &      & 11   &    &      \\
density estimation             &      & 12   &    &      \\
order statistics               &      & 13   & 10 &      \\
Markov Chain Monte Carlo       &      & 14   & 1  & 4    \\
smoothing                      &      & 16   &    &      \\
gibbs sampler                  &      & 17   &    &      \\
Bayesian inference             &      & 18   & 14 & 11   \\
efficiency                     &      & 19   &    &      \\
model selection                &      &      & 8  & 5    \\
longitudinal data              &      &      & 9  & 14   \\
measurement error              &      &      & 16 &      \\
random effects                 &      &      & 17 &      \\
variable selection             &      &      & 18 & 1    \\
empirical likelihood           &      &      & 19 & 7    \\
quantile regression            &      &      &    & 9    \\
lasso                          &      &      &    & 12   \\
sparsity                       &      &      &    & 16   \\
functional data analysis       &      &      &    & 17   \\
causal inference               &      &      &    & 18   \\ \bottomrule
\end{tabular}

\vspace{1ex}
\footnotesize 
Note: `80-90', `90-00', `00-10' and `10-20' represent the four decades \\ ``1981-1990'', ``1991-2000'', ``2001-2010'' and ``2011-2020'', respectively.
\end{table}

\begin{table}[htbp]
\centering
\small
\spacingset{0.9}
\caption{The ranks of the selected influential research fields for nodal popularity in the four decades}
\label{Beta-FeaturesTable}
\begin{tabular}{@{}lllll@{}}
\toprule
\textbf{\textbf{Nodal popularity($\beta$)-related fields}}                                        & \textbf{80-90}                      & \textbf{90-00}                      & \textbf{00-10}                        & \textbf{10-20}                      \\ \midrule
simulation                     & 1    &      &    &      \\
bootstrap                      & 2    & 1    & 2  & 4    \\
reliability                    & 3    &      &    &      \\
poisson regression             & 4    &      &    &      \\
empirical bayes                & 5    &      &    &      \\
consistency                    & 6    & 6    & 8  & 15   \\
likelihood ratio test          & 7    & 18   &    &      \\
longitudinal data              & 8    &      & 11 & 6    \\
mean squared error             & 9    &      &    &      \\
successive difference analysis & 10   &      &    &      \\
analysis of variance           & 11   &      &    &      \\
local optimality               & 12   &      &    &      \\
parameter estimation           & 13   &      &    &      \\
aligned ranks                  & 14   &      &    &      \\
normal distribution            & 15   &      &    &      \\
maximum likelihood estimation  & 16   & 15   & 19 & 10   \\
maximum likelihood             & 17   & 3    & 10 & 19   \\
growth curve                   & 18   &      &    &      \\
logistic regression            & 19   & 8    & 16 &      \\
order statistics               & 20   & 14   & 7  &      \\
nonparametric regression       &      & 2    & 3  &      \\
asymptotic normality           &      & 4    & 4  & 3    \\
EM algorithm                   &      & 5    & 5  & 2    \\
robustness                     &      & 7    & 6  & 13   \\
density estimation             &      & 9    &    &      \\
missing data                   &      & 10   & 14 & 20   \\
Gibbs sampling                 &      & 11   &    &      \\
survival analysis              &      & 12   & 13 & 14   \\
Markov Chain Monte Carlo       &      & 13   & 1  & 5    \\
smoothing                      &      & 16   &    &      \\
gibbs sampler                  &      & 17   &    &      \\
efficiency                     &      & 19   &    &      \\
Bayesian inference             &      & 20   & 12 & 9    \\
model selection                &      &      & 9  & 7    \\
measurement error              &      &      & 15 &      \\
variable selection             &      &      & 17 & 1    \\
time series                    &      &      & 18 &      \\
random effects                 &      &      & 20 &      \\
empirical likelihood           &      &      &    & 8    \\
quantile regression            &      &      &    & 11   \\
lasso                          &      &      &    & 12   \\
sparsity                       &      &      &    & 16   \\
functional data analysis       &      &      &    & 17   \\ 
causal inference               &      &      &    & 18   \\ \bottomrule
\end{tabular}

\vspace{1ex}
\footnotesize 
Note: `80-90', `90-00', `00-10' and `10-20' represent the four decades \\ ``1981-1990'', ``1991-2000'', ``2001-2010'' and ``2011-2020'', respectively.
\end{table}

Table \ref{Beta-FeaturesTable} presents the selected influential research fields associated with nodal popularity and their ranks over time. 
As nodal popularity characterizes a node’s inherent characteristics to attract connections from other nodes, in the author-citation network, it directly reflects the perceived influence and citation pull of a scholar’s publications. 
Scholars with higher popularity typically generate more inward edges and exhibit larger in-degrees.
Therefore, research fields with rising rankings indicate increasing contribution to enhancing scholars’ citation pull and collaborative influence. 
These areas are typically influential topics undergoing fast development, marked by broad methodological development, strong theoretical support, and growing interdisciplinary reach. 
For example, \textit{model selection} and \textit{variable selection} gained tremendous momentum in the 2000s with the arrival of the Big Data era  \citep{fan2008sure, Li01092012}, and have become increasingly popular with a series of influential works proposed during the early 21st century \citep{fan2001variable,fan2020statistical}.
This trajectory aligns with the results in Table \ref{Beta-FeaturesTable} that \textit{model selection} and \textit{variable selection} were not selected before 2000, yet they have become prominent as one of the most popular research fields since the beginning of the 21st century, culminating in the first-place ranking for \textit{variable selection} in 2011–2020. Similar influential popularity-related topics with rising rankings include \textit{Markov Chain Monte Carlo}, \textit{EM algorithm}, and \textit{Bayesian inference}.

Likewise, declining rankings among popularity-related research fields indicate that publications in these fields become less appealing for future citations. This pattern potentially implies a state of field maturity, where foundational frameworks are already widely established and additional publications will have diminishing benefits in terms of the connection in author-citation networks.  
For instance, \textit{analysis of variance} (ANOVA) is one of the most classic statistical topics that has been extensively formalized and disseminated, including comprehensive coverage in many standard textbooks \citep{scheffe1999analysis, casella2024statistical}. As shown in the table, while it attracted active citation engagement in the 1981–1990 period with rank 10th, it was no longer identified as a top contributor to nodal popularity after 1990. 
Another similar case is the \textit{growth curve model}, a type of generalized multivariate analysis-of-variance (GMANOVA), which ranked 18th in 1980-1990. While it remains an important topic today, scholarly attention has been shifted toward derivative research directions of GMANOVA, such as \textit{survival analysis} and \textit{longitudinal data analysis} \citep{das2008generalized}.

\section{Conclusion and Discussion}

In this paper, we propose NetworkNet, a deep neural network approach for identifying the complex nodal heterogeneity in count-valued network data with high-dimensional nodal attributes. By combining the advantages of both statistical random network models and deep learning techniques, NetworkNet can effectively model the nodal heterogeneity and simultaneously select the influential subset of nodal attributes using a specially designed DNN with two skip layers. Theoretically, we prove the consistency of estimation for nodal heterogeneity and provide the specific non-asymptotic bound for approximation error. Empirically, the dominant performances of NetworkNet in both nodal heterogeneity estimation and nodal attribute selection are shown with comprehensive simulations. By investigating author-citation networks among statisticians with the newly proposed NetworkNet, we select the influential research fields in four decades and provide several insightful findings about the evolution of scientific fields and scholarly impact. 

While our primary focus has been on count-valued data motivated by author-citation networks, the methodology can be readily extended to other network structures, including binary and continuously weighted networks, by tailoring the likelihood-based loss function. 
The compelling empirical performance of NetworkNet highlights the value of integrating statistical network modeling with deep learning techniques, and opens promising avenues for future research. For instance, extending the model to signed networks with negative edge weights introduces new challenges, particularly regarding the interpretation of selected attributes;  Applying this framework to complex economic systems (such as firm covariates in trade networks or balance sheets in interbank lending) can offer a valuable tool for investigating network formation and systemic risk. 
Furthermore, exploring the interaction effects among selected influential nodal attributes remains an unresolved challenge. 

Overall, as a general framework to model complex random network data with high-dimensional nodal attributes via neural networks, NetworkNet provides a powerful tool to capture nodal heterogeneity and select influential nodal attributes. Its flexibility and strong empirical performance make it well-suited for a wide range of interdisciplinary applications in economic, social, and biological network studies.

\spacingset{1.5}
\setlength{\bibsep}{2pt} 
\bibliography{reference}

\end{document}